\documentclass[12pt]{article}
\usepackage{latexsym}
\usepackage{epsf}

\newcommand{\andvol}[3]{{\bf #1}~(#3)~#2}

\newcommand{\PRD}[3]{Phys.~Rev.~\andvol{D#1}{#2}{#3}}

\newcommand{\PLB}[3]{Phys.~Lett.~\andvol{B#1}{#2}{#3}}

\newcommand{\hepph}[1]{\ hep-ph/#1}

\newcommand{\dfrac}[2]{\frac{\displaystyle #1}{\displaystyle #2}}

\begin{document}

\renewcommand{\thefootnote}{\fnsymbol{footnote}}

\begin{titlepage}

\begin{flushright}
VPI--IPPAP--99--02\\
hep--ph/9903362\\
March 1999
\end{flushright}

\bigskip
\bigskip
\bigskip
\bigskip

\begin{center}

\textbf{\large
Charge Assignments in Multiple--$U(1)$ Gauge Theories
}

\bigskip
\bigskip
\bigskip

\textsc{Will~LOINAZ}\footnote{%
electronic address: loinaz@alumni.princeton.edu}
and
\textsc{Tatsu~TAKEUCHI}\footnote{%
electronic address: takeuchi@vt.edu}
\\
\medskip
\textit{Institute for Particle Physics and Astrophysics\\
Physics Department, Virginia Tech, Blacksburg, VA 24061}
\\

\bigskip
\bigskip
\bigskip
\bigskip
\bigskip

\begin{abstract}
We discuss the choice of gauge field basis in 
multiple--$U(1)$ gauge theories.  
We find that there is a preferred basis, specified by the 
\textit{charge orthogonality condition}, in which the $U(1)$
gauge fields do not mix under one--loop renormalization group running.
\end{abstract}

\end{center}

\vfill

\begin{flushleft}
VPI--IPPAP--99--02\\
hep--ph/9903362\\
March 1999
\end{flushleft}

\end{titlepage}



\begin{flushleft}
\textbf{\large 1. Introduction}
\end{flushleft}
\medskip

In Refs.~\cite{LANE:95,LANE:96}, Lane and Eichten discuss
topcolor--assisted technicolor \cite{HILL:95} models
which include two $U(1)$'s as part of their gauge groups.
The charges of the fermions under the two $U(1)$'s were
restricted by anomaly cancellation conditions and
phenomenological requirements, but beyond that they
were free to move in a relatively large parameter space.
  
In a general model with multiple $U(1)$'s, including
those of Lane and Eichten, radiative corrections will mix the $U(1)$'s 
at different scales even if the gauge fields were orthogonal at the 
initial scale.
As a result, the meaning of the $U(1)$ charges assigned to the matter
fields becomes ambiguous and the phenomenology is difficult to decipher
or control.

In this letter, we argue that
for any multiple--$U(1)$ gauge theory, one can always find a
basis for the gauge fields in which this scale--dependent
mixing does not occur (at one--loop order).
In this basis, the $U(1)$ charges of the matter fields satisfy
a constraint which we call the \textit{charge orthogonality 
condition} (COC), and both the $U(1)$ fields and charges become
scale--independent concepts.

Since any multiple--$U(1)$ gauge theory is equivalent to
another in which the COC is satisfied, one can impose the
COC on the matter--field charges without loss of generality.
Not only does this simplify the study of the model
at different scales, but also places a welcome additional
restriction on the allowed charge parameter space.

\bigskip
\begin{flushleft}
\textbf{\large 2. $U(1)$ Gauge Boson Mixing}
\end{flushleft}
\medskip

Consider a theory consisting of the gauge group 
$U_1(1) \times U_2(1) \times \cdots \times U_N(1)$
coupled to fermions $\psi_a$.\footnote{Our results apply equally
to models with scalars coupled to multiple $U(1)$ fields, 
but for simplicity we restrict our attention to a model with only fermions.}
The Lagrangian one might naively write down for
such a theory is:
\[
\mathcal{L} = 
-\sum_{i=1}^N \frac{1}{4 g_i^2}F_{\mu\nu}^i F^{\mu\nu}_i
+\sum_a \bar{\psi_a}
\left( \partial_\mu - \sum_{i=1}^N Y_i^a A_\mu^i \right)
\gamma^\mu \psi_a.
\]
Here, $g_i$ is the gauge coupling of the $i$--th $U(1)$ gauge
group and $Y_i^a$ is the $i$--th $U(1)$ charge of the $a$--th
fermion.  
However, for $U(1)$ gauge fields the field strength tensors
\[    F_{\mu\nu}^i = \partial_\mu A_\nu^i - \partial_\nu A_\mu^i   \]
are by themselves gauge invariant, so terms of the form
\[    F_{\mu\nu}^i F^{\mu\nu}_j,\qquad (i\neq j)    \]
are also permitted. 
So, the most general Lagrangian allowed by the
symmetry is actually
\[
\mathcal{L}
= -\frac{1}{4} \sum_{i,j} F_{\mu\nu}^i K_i^j F^{\mu\nu}_j
+\sum_a \bar{\psi_a}
\left( \partial_\mu - \sum_{i=1}^N Y_i^a A_\mu^i \right)
\gamma^\mu \psi_a,
\]
where the $N\times N$ matrix
$K_i^j$ can be assumed to be real, symmetric, and 
positive definite.
Therefore, in general, the
$U(1)$ gauge fields will mix, and the diagonal elements
of $K_i^j$ do not have a simple interpretation as coupling
constants.

Of course, it is always possible to choose a basis which
diagonalizes $K_i^j$
and to identify its diagonal elements as the
inverse coupling constants squared in that particular basis.
However, even if $K_i^j$ is diagonalized at one scale,
renormalization reintroduces the off--diagonal 
mixing terms at other scales, and the mutually orthogonal $U(1)$ fields (and
the charges to which they couple) must be redefined at 
every new scale.  For a general $U(1)_1 \times U(1)_2 \times \cdots
\times U(1)_N$ theory (as might be embedded in models such as
topcolor--assisted technicolor) this scale--dependent mixing of
the gauge fields, the gauge couplings, and the charges of the matter fields 
will likely make the analysis of its phenomenology confusing.
The natural question arises whether it is possible to
find a basis in which $K_i^j$ is diagonal regardless of scale.  
The answer, at least at one loop, is \textit{yes}.

To construct such a basis, first diagonalize $K_i^j$
at some initial scale $\mu$:
\[
K_i^j(\mu) = 
\left[ \begin{array}{cccc} \dfrac{1}{\kappa_1^2} & & & 0 \\
                           & \dfrac{1}{\kappa_2^2} & &   \\
                           & & \ddots & \\
                           0 & & & \dfrac{1}{\kappa_N^2}
       \end{array}
\right]
\]
Next, rescale the rotated gauge fields and charges:
\begin{eqnarray*}
\frac{1}{\kappa_i}A_\mu^i & \longrightarrow & A_\mu^i, \cr
\kappa_i Y_i^a            & \longrightarrow & Y_i^a.
\end{eqnarray*}
This rescaling leaves the coupling term between 
the gauge bosons and fermions invariant 
while reducing $K_i^j(\mu)$ to a unit matrix:
\[
K_i^j(\mu) \longrightarrow \delta_i^j.
\]
In other words, we can set all the gauge coupling constants to
one by absorbing them into the definition of the charges.

Under renormalization, fermion loops induce scale--dependent mixing of the 
gauge fields which were diagonalized at scale $\mu$.  
That is, generically $K_i^j$ receives both diagonal and 
off--diagonal contributions.  
At one loop, one finds at scale $\mu'$:
\[
K_i^j(\mu')
= \delta_i^j - M_i^j \ln\left( \frac{\mu'}{\mu}\right)^2,
\]
where
\[
M_i^j 
= \frac{1}{24\pi^2}\sum_{a} {Y_i^a}{Y_j^a}
\]
in which the index $a$ runs over the \textit{chiral} fermions.

If the $U(1)$ charges satisfy the condition
\begin{equation}
\sum_{a} {Y_i^a}{Y_j^a} = 0\qquad \mbox{if $i\neq j$},
\label{COC}
\end{equation}
then $M_i^j$ is diagonal.  If it is not, 
one can always diagonalize $M_i^j$ by an orthonormal rotation 
of the gauge fields and charges which leaves both
$\sum_{i=1}^N Y_i^a A_\mu^i$ and $K_i^j(\mu)$ (the unit matrix)  
invariant.
In this new basis, 
both $K_i^j(\mu)$ and $M_i^j$ are diagonal and as a result, 
$K_i^j(\mu')$ remains diagonal for all scales $\mu'$.  
Note that the new rotated charges now satisfy Eq.~(\ref{COC}).

This discussion illustrates that it is always possible to find
a basis in which $K_i^j(\mu)$ and $M_i^j$ are simultaneously diagonalized
through \textit{rescaling} and \textit{orthonormal rotations}.
(For $K_i^j(\mu)$ and $M_i^j$ to be simultaneously diagonalizable by
orthonormal rotations alone, they must commute.)
In this basis, which is characterized by the fact that Eq.~(\ref{COC})
is satisfied, the $U(1)$ fields and charges do not mix 
under one loop running.  Hence, this should be the basis of choice
for analyzing the phenomenology of the theory.

However, starting with a general multiple--$U(1)$ theory
with arbitrary (anomaly--free) charge assignments and then
finding the preferred basis in which Eq.~(\ref{COC}) is satisfied
is tantamount to imposing Eq.~(\ref{COC}) on the charges
to begin with.
Therefore, without loss of generality, one can impose
Eq.~(\ref{COC}) as an additional condition 
to anomaly cancellations
when making the initial charge assignments.
We call Eq.~(\ref{COC}) the
\textit{charge orthogonality condition} (COC).

Since $M_i^j$ is real, symmetric, and positive semi--definite,
all of its eigenvalues are real and non--negative.  Let us denote them by:
\[
0 \le \lambda_1 \le \lambda_2 \le \cdots \le \lambda_N.  
\]
After diagonalization
\[
K_i^j(\mu')
= \delta_i^j
  \left[ 1 - \lambda_i \ln\left( \frac{\mu'}{\mu} \right)^2 \right],
\]
and we can make the identification
\[
\frac{1}{g_i^2(\mu')}
\equiv 1 - \lambda_i \ln\left( \frac{\mu'}{\mu} \right)^2.
\]

We make the following observations:
\begin{enumerate}
\item
The ratios of the $U(1)$ coupling constants $g_i(\mu)$
flow towards infrared fixed points determined by the
ratios of the eigenvalues of $M_i^j$:
\[
\dfrac{g_i^2(\mu)}{g_j^2(\mu)}\;\;
\stackrel{\mu\rightarrow 0}{\Longrightarrow}\;\;
\dfrac{\lambda_j}{\lambda_i}.
\]

\item  If the fermions are massive and decouple in the course of 
renormalization group running, 
the effective theory will change as each threshold is crossed.
Consequently, $M_i^j$ will differ across each threshold, and it will
be necessary to rediagonalize $M_i^j$ in each new effective theory.

\item
The position of the Landau pole ($\Lambda$) is determined by the
largest eigenvalue, $\lambda_N$, of $M_i^j$:
\[
\log\left( \dfrac{ \Lambda^2 }{ \mu^2 } \right)
= \dfrac{ 1 }{ \lambda_N }.
\]

\end{enumerate}

In a general basis (i.e. one in which $M_i^j$ is not diagonal) one might
naively calculate the Landau pole from the diagonal 
elements alone.  However, the largest eigenvalue of a positive semi--definite 
matrix is always at least as large as its largest diagonal element.  
Thus, the \textit{actual} Landau pole of the model is always lower than 
one would estimate neglecting gauge field mixing \cite{POPOVIC:98}.

\noindent

\bigskip
\begin{flushleft}
\textbf{\large 3. Charge Orthogonality in GUT's}
\end{flushleft}
\medskip

If the multiple $U(1)$'s under consideration are
unbroken subgroups of a simple Lie Group,
as they would be in a Grand Unified Theory (GUT),
then it is possible to show that the charge orthogonality
condition Eq.~(\ref{COC}) is automatically satisfied.
If the unbroken $U(1)$'s result from the breaking of a 
simple Lie group, the charge $Y_i^a$ of the $a$--th fermion under the 
i--th $U(1)$ is proportional to the $a$--th diagonal element of 
the $i$--th unbroken diagonal generator of the simple Lie group
in some irreducible representation. 
The diagonal generators $H_i$ $(i=1,2,\cdots,N)$ of a simple Lie group 
constitute the Cartan subalgebra of the generating Lie algebra 
and can always be normalized to the form
\[
\mathrm{tr}(H_i H_j) = \delta_{ij} k_D
\]
where $k_D$ is a representation--dependent constant \cite{GEORGI}.  
Since  $\sum_{a}{Y_i^a Y_j^a}$ is proportional
to ${\rm tr}(H_i H_j)$, it immediately follows that, for $U(1)$'s arising
from the breaking of simple Lie groups, the COC is always satisfied.

Of course, if any of the fermions decouple in the course of 
renormalization group running before the $U(1)$'s themselves break,
then a re--diagonalization of the fields and charges would be
necessary every time a particle threshold is crossed even for
GUT's.

\bigskip
\begin{flushleft}
\textbf{\large 4. Application to Topcolor--Assisted Technicolor Models}
\end{flushleft}
\medskip

\renewcommand{\arraystretch}{1.2}

\begin{table}[ht]
\begin{center}
\begin{tabular}{|c|c|c|c|}
\hline
Particle & $Y_1$ & $Y_2$ & $Q=T_3 + Y_1 + Y_2$ \\
\hline\hline
$q_L^l$   & 0 & $\frac{1}{6}$ & $\frac{2}{3}, -\frac{1}{3}$ \\
\hline
$c_R,u_R$ & 0 & $\frac{2}{3}$ & $\phantom{+}\frac{2}{3}$ \\
\hline
$d_R,s_R$ & 0 & $-\frac{1}{3}$ & $-\frac{1}{3}$ \\
\hline
$q_L^h$ & $\frac{1}{6}$ & 0 & $\frac{2}{3}, -\frac{1}{3}$ \\
\hline
$t_R$   & $\frac{2}{3}$ & 0 & $\frac{2}{3}$ \\
\hline
$b_R$   & $-\frac{1}{3}$ & 0 & $-\frac{1}{3}$ \\
\hline
$T_L^l$ & $x_1$ & $x_2$               & $\pm\frac{1}{2} + x_1 + x_2$ \\
\hline
$U_R^l$ & $x_1$ & $x_2+\frac{1}{2}$   & $\phantom{+}\frac{1}{2} + x_1 + x_2$ \\
\hline
$D_R^l$ & $x_1$ & $x_2-\frac{1}{2}$   & $-\frac{1}{2} + x_1 + x_2$ \\
\hline
$T_L^t$ & $y_1$ & $y_2$               & $\pm \frac{1}{2} + y_1 + y_2$ \\
\hline
$U_R^t$ & $y_1+\frac{1}{2}$ & $y_2$   & $\phantom{+}\frac{1}{2} + y_1 + y_2$ \\
\hline
$D_R^t$ & $y_1+\frac{1}{2}$ & $y_2-1$ & $-\frac{1}{2} + y_1 + y_2$ \\
\hline
$T_L^b$ & $z_1$             & $z_2$   & $\pm\frac{1}{2} + z_1 + z_2$ \\
\hline
$U_R^b$ & $z_1-\frac{1}{2}$ & $z_2+1$ & $\phantom{+}\frac{1}{2} + z_1 + z_2$ \\
\hline
$D_R^b$ & $z_1-\frac{1}{2}$ & $z_2$   & $-\frac{1}{2} + z_1 + z_2$ \\
\hline
\end{tabular}
\caption{Quark and technifermion hypercharges and electromagnetic charges 
in the model of Ref.~\cite{LANE:95}.  Leptons are not shown.}
\end{center}
\end{table}

To see the consequence of the COC in model--building, 
we consider its application in the context of 
topcolor--assisted technicolor models \cite{HILL:95}. 
As a concrete example, we consider the model proposed by Lane and Eichten
in Ref.~\cite{LANE:95}, which has two U(1) factors in its gauge group.
The particle content and charge assignments of the model are shown in Table~1,
where the $U(1)$ charges of technifermions are expressed in terms
of six parameters $x_{1,2}$, $y_{1,2}$, and $z_{1,2}$.
(Note that the Standard Model particles are charged under only one of 
the $U(1)$'s.)

Anomaly cancellation requirements lead to the
following five equations for the six unknowns:
\begin{eqnarray*}
0 & = & x_1 + y_1 + z_1,  \cr
0 & = & x_2 + y_2 + z_2,  \cr
0 & = & x_1 \left( y_1 - z_1 + \frac{1}{2} \right), \cr
0 & = & x_2 \left( y_2 - z_2 - \frac{1}{2} \right), \cr
0 & = & x_1 \left( y_2 - z_2 - \frac{1}{2} \right) + 
        x_2 \left( y_1 - z_1 + \frac{1}{2} \right).
\end{eqnarray*}
There are two classes of solutions to these
equations: the first with $x_1 = x_2 = 0$, and the other
with $(y_1 - z_1 + \frac{1}{2}) = (y_2 - z_2 - \frac{1}{2}) = 0$.
Each of these classes has two free parameters remaining.
The COC provides an additional constraint
\[
4 (x_1 x_2 + y_1 y_2 + z_1 z_2) 
- \left( y_1 - z_1 + \frac{1}{2} \right) 
+ \left( y_2 - z_2 - \frac{1}{2} \right) 
= 0,
\]
which reduces the number of free parameters to one in
each class.

Instead of applying the COC, however, Lane and Eichten require the
existence of the four--technifermion operator
\[
\bar{T}_L^l \gamma^\mu T_L^t\,
\bar{D}_R^t \gamma_\mu D_R^b
\]
to provide mixing between the third and first two fermion generations.
This leads to two additional constraints
\begin{eqnarray*}
0 & = & x_1 - z_1 + 1, \cr
0 & = & x_2 - z_2 - 1,
\end{eqnarray*}
which are sufficient to fix one solution in each of the two classes.
However, these conditions, and hence the two solutions, are incompatible 
with the COC.

This means that the charge assignments of Ref.~\cite{LANE:95} characterize 
a basis for the $U(1)$ fields which is \textit{not} charge--orthogonal, 
and in this basis the $U(1)$ fields will necessarily mix with a 
scale--dependent mixing angle under RG running. 
To avoid this, one must go to the preferred basis in which the COC
is satisfied, but in that basis the Standard Model particles 
will carry two $U(1)$ charges instead of just one.
We are therefore forced to consider more general
charge assignments, 
such as those considered by Lane in Ref.~\cite{LANE:96}.

The model constructed by Lane in Ref.~\cite{LANE:96}
is similar to that of Ref.~\cite{LANE:95}, but the $U(1)$ charge assignments
of all the particles, including those in the Standard Model, 
are kept generic resulting in 26 parameters which
must be specified.
Anomaly cancellations provide a set of
five linear and three cubic equations.
Thirteen additional linear equations are imposed from phenomenological
considerations.
With 21 conditions on 26 parameters, any solution will have
at least five free parameters.
The COC will provide an additional quadratic equation 
which will again reduce the number of free parameters by
removing the (unphysical) 
rescaling and rotational degree of freedom in the 
$U(1)$ field and charge definitions.

\bigskip
\begin{flushleft}
\textbf{\large 5. Conclusion}
\end{flushleft}
\medskip

We have argued that in multiple--$U(1)$ gauge theories,
there exists a particular choice of basis for the $U(1)$ fields 
which is to be preferred over other rescaling and orthonormal rotations 
of the $U(1)$ fields.  
In this basis, characterized by the fact that the COC is satisfied, 
the $U(1)$ gauge bosons do not mix under one--loop renormalization group 
running.
Choosing this basis in model--building avoids 
the ambiguity in charge definitions associated with rescaling and
rotations of the gauge fields 
and expedites study of the models at different scales.

Since a theory with arbitrary $U(1)$ charge assignments can always
be mapped onto another in which the COC is satisfied
with a simple change of basis,
one can impose the COC as an additional constraint
on the charges without loss of generality.
This will reduce the dimension of the available charge
parameter space and simplify the analysis of models considerably.

\bigskip
\begin{flushleft}
\textbf{\large Acknowledgement}
\end{flushleft}
\medskip

We thank T.~J.~Newman for helpful conversations and for careful
reading of this manuscript.
This work was supported in part (W.L.) by the U.S. Department of Energy,
grant DE--FG05--92--ER40709, Task~A.


\end{document}